\documentclass[aps,prl,preprintnumbers,amsmath,amssymb,latexsym,array,enumerate,letter,twocolumn,superscriptaddress,showpacs]{revtex4-1}

\pdfoutput=1

 \usepackage{titlesec}
   \titleformat{\section}[runin]
   {\normalfont\bfseries}{\quad\thesection.}{0.5em}{} [.]
  \titlespacing*{\section}{0pt}{0.2\baselineskip}{\baselineskip}

\usepackage{graphicx,color}
\usepackage{dcolumn}

\usepackage{caption}
\usepackage{subcaption}
\newcommand{\Tr}{\ensuremath{\operatorname{Tr}}}

\begin{document}
\title{\color{blue}\Large  Quark matter may not be strange}

\author{Bob Holdom}
\email{bob.holdom@utoronto.ca}
\author{Jing Ren}
\email{jren@physics.utoronto.ca}
\author{Chen Zhang}
\email{czhang@physics.utoronto.ca}
\affiliation{Department of Physics, University of Toronto, Toronto, Ontario, Canada  M5S1A7}

\begin{abstract}

If quark matter is energetically favored over nuclear matter at zero temperature and pressure then it has long been expected to take the form of strange quark matter (SQM), with comparable amounts of $u$, $d$, $s$ quarks. The possibility of quark matter with only $u$, $d$ quarks ($ud$QM) is usually dismissed because of the observed stability of ordinary nuclei. However we find that $ud$QM generally has lower bulk energy per baryon than normal nuclei and SQM. This emerges in a phenomenological model that describes the spectra of the lightest pseudoscalar and scalar meson nonets. Taking into account the finite size effects, $ud$QM can be the ground state of baryonic matter only for baryon number $A>A_\textrm{min}$ with $A_\textrm{min}\gtrsim 300$. This ensures the stability of ordinary nuclei and points to a new form of stable matter just beyond the periodic table.

\end{abstract}
\maketitle

\section{Introduction}

Hadronic matter is usually thought to be the ground state of baryonic matter (matter with net baryon number) at zero temperature and pressure. Then quark matter only becomes energetically favorable in an environment like at a heavy ion collider or deep inside the neutron star. 
However as proposed by Witten~\cite{Witten} (with some relation to earlier work~\cite{Itoh:1970uw,Bodmer:1971we,Terazawa:1979hq,Chin:1979yb}), quark matter with comparable numbers of $u, d, s$, also called strange quark matter (SQM), might be the ground state of baryonic matter, with the energy per baryon $\varepsilon\equiv E/A$ even smaller than 930\,MeV for the most stable nuclei ${}^{56}$Fe. 

With the lack of a first-principles understanding of the strong dynamics, the MIT bag model~\cite{Chodos:1974je} has long been used as a simple approximation to describe quark matter. In this model constituent quark masses vanish inside the bag, and SQM is found to reach lower energy than quark matter with only $u, d$ quarks ($ud$QM). 
If SQM is the ground state down to some baryon number $A_{\rm min}$, as long as the transition of ordinary heavy nuclei with $A>A_{\rm min}$ to SQM needs a simultaneous conversion of a sufficiently large number of down quarks to strange quarks, the conversion rate can be negligibly small \cite{Jaffe}. A faster catastrophic conversion could occur if the ground state was instead $ud$QM.
So quite often the energy per baryon is required to satisfy $\varepsilon_\textrm{SQM}\lesssim 930\,\textrm{MeV}\lesssim \varepsilon_\textrm{$ud$QM}$~\cite{Jaffe,2015Compact}.

On the other hand, since the periodic table of elements ends for $A\gtrsim300$, this catastrophe can be avoided if $A_\textrm{min}\gtrsim 300$ for $ud$QM. 
It is also recognized that the bag model may not adequately model the feedback of a dense quark gas on the QCD vacuum. How the $u$, $d$, $s$ constituent quark masses respond to the gas should account for the fact that flavor symmetry is badly broken in QCD. 
This can be realized in a quark-meson model by incorporating in the meson potential the flavor breaking effects originating in the current quark masses. 
Through the Yukawa term the quark densities drive the scalar fields away from their vacuum values. The shape of the potential will then be important to determine the preferred form of quark matter.
This effect has already been seen in NJL and quark-meson models~\cite{Buballa:1998pr,PWang01,Ratti:2002sh,Buballa:2003qv,OsipovSM}. These are studies in the bulk limit and they tend to find that $ud$QM has lower $\varepsilon$ than SQM with the conclusion that neither is stable.

The possibility that $ud$QM is actually the ground state of baryonic matter has been ignored in the literature, but it shall be our focus in this letter.
With an effective theory for only the scalar and pseudoscalar nonets of the sub-GeV mesons with Yukawa coupling to quarks, we demonstrate a robust connection between the QCD spectrum and the conditions for a $ud$QM ground state in the bulk, i.e. $\varepsilon_\textrm{$ud$QM}\lesssim 930\,\textrm{MeV}$ and $\varepsilon_\textrm{$ud$QM}< \varepsilon_\textrm{SQM}$. We shall also show that surface effects are of a size that can ensure that $A_\textrm{min}\gtrsim 300$ by numerically solving the scalar field equation of motion. This points to the intriguing possibility that a new form of stable matter consisting only of $u,d$ quarks might exist not far beyond the end of the periodic table. 

\section{The meson model\label{sec_mmodel}}

Here we study an effective theory describing the mass spectra and some decay rates of the scalar and pseudoscalar nonets of the sub-GeV mesons. The QCD degrees of freedom not represented by these mesons are assumed to be integrated out and encoded in the parameters of the phenomenological meson potential. We view our description as dual to one that contains vector mesons \cite{Holdom:1996fj}. With the parameters determined from data, we can then extrapolate from the vacuum field values to the smaller field values of interest for quark matter.

We find that a linear sigma model provides an adequate description without higher dimensional terms,
\begin{eqnarray}\label{eq:LSM}
  \mathcal{L}_m &=& \Tr \left( \partial_\mu \Phi^\dagger \partial^\mu
    \Phi \right)
  - V,\quad V=V_{\text{inv}} + V_b
.\end{eqnarray}
$ \Phi = T_a \left(\sigma_a + i \pi_a\right)$ is the meson field and $T_a = \lambda_a/2$ ($a=0,\ldots, 8$) denotes the nine generators of the flavor $U(3)$ with $\Tr (T_a T_b) = \delta_{ab}/2 $. 
$V_{\text{inv}}$ is chirally invariant, 
\begin{equation}
\begin{aligned}
V_{\text{inv}}&=\lambda_1\left(\Tr\Phi^\dagger\Phi\right)^2 +\lambda_2\Tr \left((\Phi^\dagger\Phi)^2\right) \\
 &+m^2\Tr\left(\Phi^\dagger\Phi\right) -c\left(\det\Phi+h.c. \right).
 \end{aligned}
 \end{equation}
The $c$ term is generated by the 't Hooft operator. Boundedness from below requires that $\lambda_1+ \lambda_2/2>0$. For there to be spontaneous symmetry breaking in the absence of $V_b$ requires that $8m^2(3\lambda_1+\lambda_2)<c^2$. 
 
$V_b=\sum_{i=1}^{8}V_{bi}$ describes the explicit $SU(3)$ flavor breaking by incorporating the current quark mass matrix ${\cal M}=\text{diag}(m_{u0},m_{d0}, m_{s0})$.
\begin{eqnarray}
\label{L-chi-1}
   V_{b1}&=&b_{1} \Tr\left(\Phi^\dagger{\cal M} +h.c. \right),
   \nonumber \\
   V_{b2}&=&b_{2}\epsilon_{ijk}\epsilon_{mnl}
   {\cal M}_{im}\Phi_{jn}\Phi_{kl}+h.c.\,,
   \nonumber \\
   V_{b3}&=&b_{3}\Tr
   \left(\Phi^\dagger\Phi\Phi^\dagger{\cal M}\right)+h.c. \,,
   \nonumber \\
   V_{b4}&=&b_{4}\Tr\left(\Phi^\dagger\Phi\right)
   \Tr\left(\Phi^\dagger{\cal M}\right)+h.c.\,,
   \nonumber \\
   V_{b5}&=&b_{5}\Tr\left(\Phi^\dagger{\cal M}
   \Phi^\dagger{\cal M}\right)+h.c.\,,
   \nonumber \\
   V_{b6}&=&b_{6}\Tr\left(\Phi\Phi^\dagger{\cal M}
   {\cal M}^\dagger +\Phi^\dagger\Phi{\cal M}^\dagger{\cal M}\right),
   \nonumber \\
   V_{b7}&=&b_{7}\left(\Tr\Phi^\dagger{\cal M}
   + h.c.\right)^2,
   \nonumber \\
   V_{b8}&=&b_{8}\left(\Tr\Phi^\dagger{\cal M}
   - h.c.\right)^2.
\end{eqnarray}
Other possible terms have been eliminated by a field redefinition \cite{OsipovMass}. We adopt $m_{s0}=94\,$MeV and $m_{ud0}=3.4\,$MeV~\cite{pdg2016}. 
This general set of terms is successful at describing the lightest scalar and pseudoscalar nonets, with all masses below 1 GeV, which is typically not possible when keeping the $V_{b1}$ term only~\cite{Tornqvist99, Lenaghan00, Parganlija13}. The size of the $b_i$ coefficients are made more meaningful by normalizing w.r.t.~the estimates of Naive Dimensional Analysis (NDA)~\cite{NDA} to obtain dimensionless NDA couplings, 
\begin{eqnarray}
\bar{\lambda}_{1, 2}&=&\frac{f^2_{\pi}}{\Lambda^2}\lambda_{1, 2}, \,\, 
\bar{m}^2 =\frac{1}{\Lambda^2} m^2,\,\,
\bar{c}=\frac{f_{\pi}}{\Lambda^2}c,\,\, 
\bar{b}_{1}=\frac{1}{  f_{\pi}\Lambda} b_1, \nonumber\\
\bar{b}_{2}&=&\frac{1}{  \Lambda} b_2,\,\, 
\bar{b}_{3, 4}=\frac{f_{\pi}}{\Lambda} b_{3, 4},\,\, 
\bar{b}_{5-8}=b_{5-8}.
\label{NDA_coeff}
\end{eqnarray}
$f_\pi$ is the pion decay constant and $\Lambda=4\pi f_\pi$ is an effective cutoff.

In the meson model chiral symmetry breaking of QCD is realized by the non-zero vacuum expectation values of the neutral scalar meson fields at the potential minimum, $\langle \Phi \rangle=T_0v_{0}+T_8 v_{8} = \frac{1}{2} \text{diag}(v_n, v_n, {\sqrt{2}v_s})$, where
we use the non-strange and strange flavor basis: $\sigma_n=\frac{\sqrt2}{\sqrt3}\sigma_0+\frac{1}{\sqrt3}\sigma_8$, $\sigma_s=\frac{1}{\sqrt3}\sigma_0-\frac{\sqrt2}{\sqrt3}\sigma_8$.
The deformation by $\mathcal{M}$ naturally implies an $SU(3)$ breaking vacuum $v_n\neq\sqrt{2}v_s$.
A standard gauging of the model then leads to $v_n =  f_{\pi}=92\,$MeV, $v_s = \sqrt{2}f_K-f_\pi/\sqrt{2}=90.5\,$MeV~\cite{pdg2016,Lenaghan00}. 

The mass spectra for the scalar and pseudoscalar nonets are derived by $\mathbb{M}^2_{s,ab}=\partial^2 V/\partial \sigma_a \partial \sigma_b$ and $\mathbb{M}^2_{p,ab}=\partial^2 V/\partial \pi_a \partial \pi_b$. 
With isospin symmetry the eight independent masses are
$m^2_{a_0}= \mathbb{M}^2_{s,11}$, $m^2_{\kappa}= \mathbb{M}^2_{s,44}$, $m^2_{\pi}=\mathbb{M}^2_{p,11}$, $m^2_{K}= \mathbb{M}^2_{p,44}$ and $m^2_{\sigma},m^2_{f_0},m^2_{\eta},m^2_{\eta'}$ after diagonalizing the $(0,8)$ sectors.
The rotations are defined as: $\sigma_0=\cos\theta_s \sigma - \sin\theta_s f_0$, $\sigma_8=\sin\theta_s \sigma + \cos\theta_s f_0$ and $\pi_0=\cos\theta_p \eta' - \sin\theta_p \eta$, $\pi_8=\sin\theta_p \eta' + \cos\theta_p \eta$.  

We solve the 12 free parameters  ($\lambda_1,\, \lambda_2, \,c,\,m^2$, $\,b_1, \,...,\,b_8$) in (\ref{eq:LSM}) in terms of two decay constants, eight meson masses and two mixing angles.
$\theta_p$ is related to the diphoton radiative decay widths of $\eta', \,\eta$ and the strong decay widths of $a_0, \kappa$. $\theta_s$ needs to fit the small and large $\pi\pi$ widths of $f_0$ and $\sigma$ respectively, which implies that the $\sigma$ meson is quite close to the non-strange direction. 

\begin{table}[h]
\begin{center}
\caption{The NDA couplings for benchmarks}
\begin{ruledtabular}
\begin{tabular}{ccc c c c c}
     &  $\bar\lambda_1$ 
     & {$\bar\lambda_2$}
     & {$\bar{m}^2$}
     &{$\bar{c}$}
     & {$\bar{b}_1$}
     & {$\bar{b}_2$}
\\
&&&&
\\[-3.5mm]
\hline

\\
&&&&&
\\[-6.5mm]

\,\,Set 1\,\, &\,\,$-0.06$\,\,& 0.33& \,\,$-0.13$\,\, & 0.33& $-4.4$& 0.19 

\\
&&&&
\\[-3.5mm]

Set 2 &0.04& \,\,0.16\,\, &0.05& \,\,0.27\,\, & \,\,$-1.6$\,\, & \,\,$-0.14$\,\,

\\
&&&&
\\[-3.5mm]
\hline 

\\
&&&&
\\[-6.5mm]

     &{$\bar{b}_3$} 
     & {$\bar{b}_4$}
     &{$\bar{b}_5$}
     &{$\bar{b}_6$}
     &{$\bar{b}_7$}
     &{$\bar{b}_8$} 
\\
&&&&
\\[-3.5mm]
\hline

\\
&&&&&
\\[-6.5mm]

Set 1 & $-4.2$ & 2.5& $-3.0$& 50 & 1.4 & 4.7

\\
&&&&
\\[-3.5mm]

Set 2 & $-0.18$ & 0.09 & 4.0 & 5.2 & $-3.9$& $-5.5$
\\

\end{tabular}
\end{ruledtabular}
\label{table_coupling}
\end{center}  \
\end{table}\vspace{-0ex}

\begin{table}[h]
\vspace{-4ex}
\begin{center}
\caption{The meson masses (in MeV), mixing angles, and decay widths (in MeV, keV for scalar, pseudoscalar).}
\begin{ruledtabular}
\begin{tabular}{c c c c c c}
     &  $\,\,\,m_\pi\,\,\,$ &  $\,\,\,m_K\,\,\,$ & $m_\eta$ &  {$m_\eta'$} & $\theta_p$
\\
&&&&&
\\[-3.5mm]
\hline

\\
&&&&&
\\[-6.5mm]

Exp & 138 & 496 & 548 & $958$ & NA

\\
&&&&&
\\[-3.5mm]

\,\,Set 1\,\, & 138 & 496 &548 & 958 & $-15.0^\circ$ 
\\
&&&&&
\\[-3.5mm]
Set 2 & 148 & 454 & 569 & 922 & $-10.8^\circ$ 
\\
&&&&&
\\[-3.5mm]
\hline

     &  {$m_{a_0}$}  &  {$m_{\kappa}$}  &{$m_\sigma$}  &  {$m_{f_0}$} & $\theta_s$

\\
&&&&&
\\[-3.2mm]
\hline

\\
&&&&&
\\[-6.5mm]

Exp & $980\pm 20$ & 700-900 & 400-550 & $990\pm 20$ & NA

\\
&&&&&
\\[-3.5mm]

Set 1 & 980 & 900 & 555 & 990 & $31.5^\circ$ 
\\
&&&&&
\\[-3.5mm]
Set 2 & 887 & 916 & 555 & 955 & $21.7^\circ$

\\
&&&&&
\\[-3.5mm]
\hline

    & $\Gamma_{\eta\to\gamma \gamma}$ & $\Gamma_{\eta'\to\gamma \gamma}$ & $\,\Gamma_{\sigma\to\pi\pi}\,$ &$\,\Gamma_{\kappa\to K\pi}\,$ &

\\
&&&&&
\\[-3.5mm]
\hline

\\
&&&&&
\\[-6.5mm]

Exp & 0.52-0.54 & 4.2-4.5 &400-700 &$\sim 500$ &

\\
&&&&&
\\[-3.5mm]

Set 1 & 0.59 & 4.90 & 442 & 451 &

\\
&&&&&
\\[-3.5mm]

Set 2 & 0.54 & 4.87 & 422 & 537 &

\\
&&&&&
\\[-3.5mm]
\hline
  & $\,\Gamma_{f_0\to\pi\pi}$ & $R_{f_0}$ & $\,\Gamma_{a_0\to\eta\pi}$ & $R_{a_0}$ &
\\
&&&&&
\\[-3.2mm]
\hline

\\
&&&&&
\\[-6.5mm]

Exp  &10-100 & 3.8-4.7 &50-100 & 1.2-1.6 &
\\
&&&&&
\\[-3.5mm]

Set 1& 11 & 4.3 & 37.4 & 2.4 &
\\
&&&&&
\\[-3.5mm]

Set 2& 20 & 4.0 & 52.0 & 1.2 &

\\
\end{tabular}
\end{ruledtabular}
\label{table_Mass}
\end{center}\vspace{-4ex}
\end{table}
 
Table~\ref{table_coupling} presents two benchmarks for the meson model. The parameters of set 1 are chosen to give a good fit to the data, however this leads to a rather large value for the NDA coupling $\bar{b}_6$. Given the theoretical uncertainties associated with the neglected higher dimensional terms, allowing the masses and decay widths to depart from the experimental values could be more sensible. An example with up to 10\% departures gives the smaller NDA couplings of set 2. 

Table~\ref{table_Mass} compares the experimental values~\cite{pdg2016} with the results of the two benchmarks, including predictions for some decay widths.  
The $f_0, a_0$ widths have large KK threshold corrections and so for these Flatt\'e~\cite{Flattee} rather than Breit-Wigner widths are used. In these cases we also compare to ratios $R_{f_0}$, $R_{a_0}$ that involve the strange and non-strange amplitudes~\cite{Bugg,OsipovDW}. We have checked that turning on explicit isospin breaking ($m_{u0}\neq m_{d0}$) has negligible impact on this study. But it does turn on the $\pi^0-\eta(\eta')$ mixing angles, $\epsilon$ and $\epsilon'$. $\epsilon$ is found roughly consistent with experiments~\cite{Osipov:2015lva, Kroll:2004rs}, while $\epsilon'$ can be compared to future measurements.

\section{Quark matter in general}
\label{sec_qm}

Now we can employ the meson model to study quark matter. Quark matter can become energetically favorable due to the reduction of the constituent quark masses in the presence of the quark densities. QCD confinement on the other hand prevents net color charge from appearing over large volumes. We suppose that these residual QCD effects on the energy per baryon are minor, similar to the way they are minor for the constituent quark model description of much of the QCD spectrum.

With the Yukawa coupling to quarks, $\mathcal{L}_y=-2g\bar{\psi}\Phi\psi$,
the equations of motion for the spherically symmetric meson fields of interest are \cite{Lee:1986tr,Bahcall:1989ff}
\begin{equation}
\begin{aligned}
 \nabla ^2\sigma_n(r)&= \frac{\partial V}{ \partial \sigma_n}+ g\sum_{i=u,d}\langle \bar{\psi_i}\psi_i \rangle, \\ 
 \nabla ^2\sigma_s(r)&= \frac{\partial V}{ \partial \sigma_s}+ \sqrt{2}g\langle \bar{\psi_s}\psi_s \rangle.
 \label{qm_eom}
\end{aligned}
\end{equation}
$\nabla ^2=\frac{d^2}{dr^2} + \frac{2}{r} \frac{d}{dr} $ and there are $N_C=3$ colours of quarks.
The quark gas is described by the Fermi momentum for each flavor $p_{Fi}=p_{F} {f_i}^{1/3}$ where the quark fractions are $f_i=n_i/(N_Cn_A)$, $p_{F}=(3\pi^2 n_A)^{1/3}$ and $n_A$ is the baryon number density. The $r$ dependence of these quantities is determined by the equations of hydrostatic equilibrium.

The forces driving the field values are from the scalar potential and the quark gas densities $\langle\bar{\psi_i}\psi_i\rangle=\frac{2N_C}{(2\pi)^3} \int_0^{p_{Fi}}  d^3p\ m_i/\sqrt{p^2+m_i^{2}}$. In the interior the quark masses $m_{u,d}(r)=g \sigma_n(r)+m_{ud0}$ and $m_{s}(r)=\sqrt{2} g \sigma_s(r)+m_{s0}$ become smaller than the vacuum values $m_{udv}$ and $m_{sv}$. The radius $R$ of the bound state is defined where $\sigma_i(r)$ and $p_{Fi}(r)$ quickly approach their vacuum values. 

Electrons play a minor role for any $A$, and they need not be contained when $R$ becomes smaller than the electron Compton wavelength, i.e. $A\lesssim 10^7$~\cite{Jaffe}. 
The quark, scalar and Coulomb energy densities are~\cite{Jaffe,Heiselberg:1993dc}
\begin{eqnarray}
\rho_{\psi}&=&\sum_{i=u,d,s} \frac{2 N_C}{(2\pi)^3}\int_0^{p_{Fi}} d^3p {\sqrt{p^2+m_i^{2}}},\nonumber\\
\rho_{\phi}&=&\Delta V+\frac{1}{2}\sum_{i=n,s}(\nabla \sigma_i)^2,\quad
\rho_Z=\frac{1}{2}\sqrt{\alpha}V_C \,n_Z\,.
\end{eqnarray}
$\Delta V$ is the potential energy w.r.t.~the vacuum. 
$n_Z=\frac{2}{3}n_u-\frac{1}{3}(n_d+n_s)$ is the charge density, $V_C$ is the electrostatic potential and $\alpha=1/137$.
The flavor composition of the quark gas and the radius $R$ can be determined by minimizing the energy of the bound state $E=\int_0^R d^3r (\rho_{\psi}+\rho_{\phi}+\rho_Z)$~\cite{Berger:1986ps}. 

\section{Quark matter in the bulk limit}
\label{subsec_bulk}

At large $A$, finite size effects can be ignored and then both the meson fields and quark densities can be taken to be spatially constant.  From (\ref{qm_eom}) and for given $(p_F,\,f_i)$ the meson fields take values where the two forces balance. Among these force balancing points we can find the values of $(\bar{p}_F,\,\bar{f}_i)$ that minimize the energy per baryon $\varepsilon=(\rho_{\psi}+\rho_{\phi}+\rho_Z)/n_A$, 
with the uniform charge density $\rho_Z=\frac{4\pi}{5}\alpha R^2 n_Z^2$. 
The flavor composition $\bar{f}_i$ is driven to charge neutrality in the large $A$ limit to avoid the dominance of $\rho_Z$.

\begin{figure}[t]
       \includegraphics[width=8.9cm]{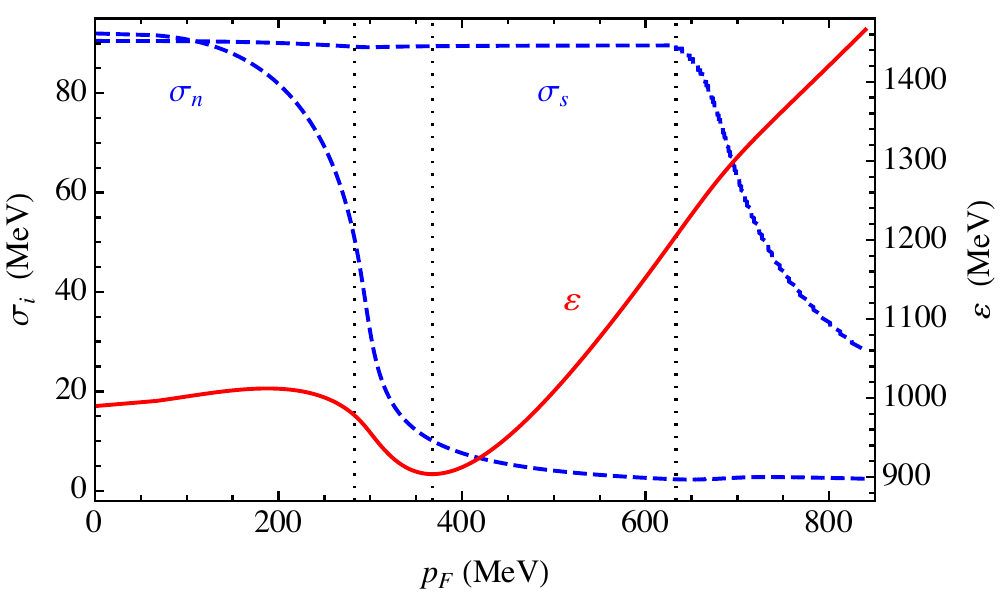}  
 \caption{The field values $\sigma_n$, $\sigma_s$ (blue dashed, left axis)  and the energy per baryon number $\varepsilon$ (red solid, right axis) in the bulk limit. The vertical lines denote the values of $p_F^{(n)}$, $\bar{p}_F$ and $p_F^{(s)}$.}
 \label{EperQ}\vspace{-3ex}
  \end{figure}

Fig.~\ref{EperQ} presents the field values and the energy per baryon as functions of $p_F$, after minimization w.r.t.~the $f_i$, for the Set 1 benchmark with $m_{udv}=330$~MeV (which implies $g=3.55$ and $m_{sv}=548$~MeV). 
The minimum energy per baryon is $\bar\varepsilon=903.6$ (905.6) MeV at $\bar{p}_{F} = 367.8$ (368.5) MeV with $\bar{f}_s\approx0$ for Set 1 (Set 2). For both sets $ud$QM is the ground state of baryonic matter in the bulk.  

As $p_F$ increases from small values the fields move away from the vacuum along the least steep direction, which is a valley oriented close to the $\sigma_n$ direction.
$\sigma_n$ drops rapidly at $p_F^{(n)}$ and at $\bar{p}_F$ the minimal energy per baryon $\bar\varepsilon$ is reached. $\bar{p}_F$ can be estimated by minimizing the relativistic quark and potential energies $\varepsilon\approx\frac{3}{4}N_C p_F \chi+3\pi^2\Delta V_n/p_F^3$ w.r.t.~$p_F$ only. $\chi=\sum_i f_i^{4/3}$ and $\Delta V_n$ is the potential difference along the valley. This gives $\bar\varepsilon\approx N_C \bar\chi\, \bar{p}_F$ and $\bar{p}_{F}^4\approx12 \pi^2\Delta V_n/(N_C\bar{\chi})$, with only $u$ and $d$ quarks contributing in $\bar\chi$. $f_s$ will finally turn on for $p_F\gtrsim p_F^{(s)}$ when it is energetically favorable to produce strange quarks (that may or may not be relativistic).

\begin{figure}[!h]
       \includegraphics[width=8cm]{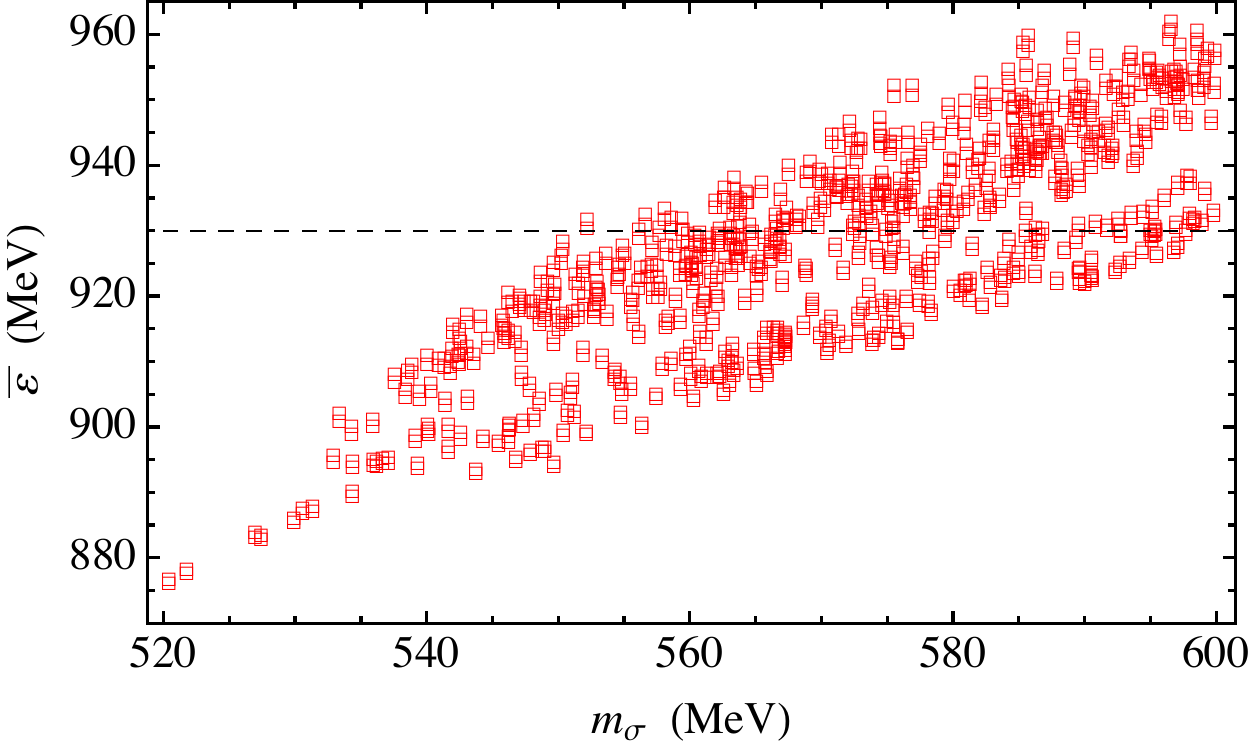}  
 \caption{$\bar\varepsilon$ vs $m_\sigma$ from the parameter scan.}
 \label{scan}\vspace{-1ex}
  \end{figure}

Our conclusion regarding $ud$QM relies on the features that $p_F^{(n)}\lesssim\bar{p}_F\lesssim p_F^{(s)}$ and $\bar\varepsilon\lesssim 930$ MeV. 
These quantities can be estimated with a parameter scan of the meson model along with $m_{udv}\approx330$-$360$~MeV. The scan is constrained to be no more than about 10\% outside the experimental ranges and with NDA coupling magnitudes less than 15. We find the ranges $p_F^{(n)}\approx280$-$305$~MeV, $\bar{p}_F\approx355$-$395$~MeV, $p_F^{(s)}\gtrsim550$~MeV and $\bar\varepsilon\approx875$-$960$~MeV.
As an example of sensitivity to the lightest meson masses, Fig.~\ref{scan} shows a $\bar\varepsilon$ vs $m_\sigma$ projection of the parameter space where we see that realistic values of $m_\sigma$ favor stable $ud$QM.

\section{Determination of $A_\textrm{min}$ for $ud$QM}
At smaller $A$ we need to include finite size effects
and the Coulomb energy contribution. 
We adopt the approximation that the values of $p_F$ and $f_i$ are constants, nonvanishing only for $r<R$, which has been found to give a good approximation for the binding energy~\cite{MW}. For each $A$ we solve for the profile of the field $\sigma_n(r)$ moving along the valley using (\ref{qm_eom}) and find the configuration, including the radius $R$, that minimizes the energy $E$.

\begin{figure}[t]
\includegraphics[width=8cm]{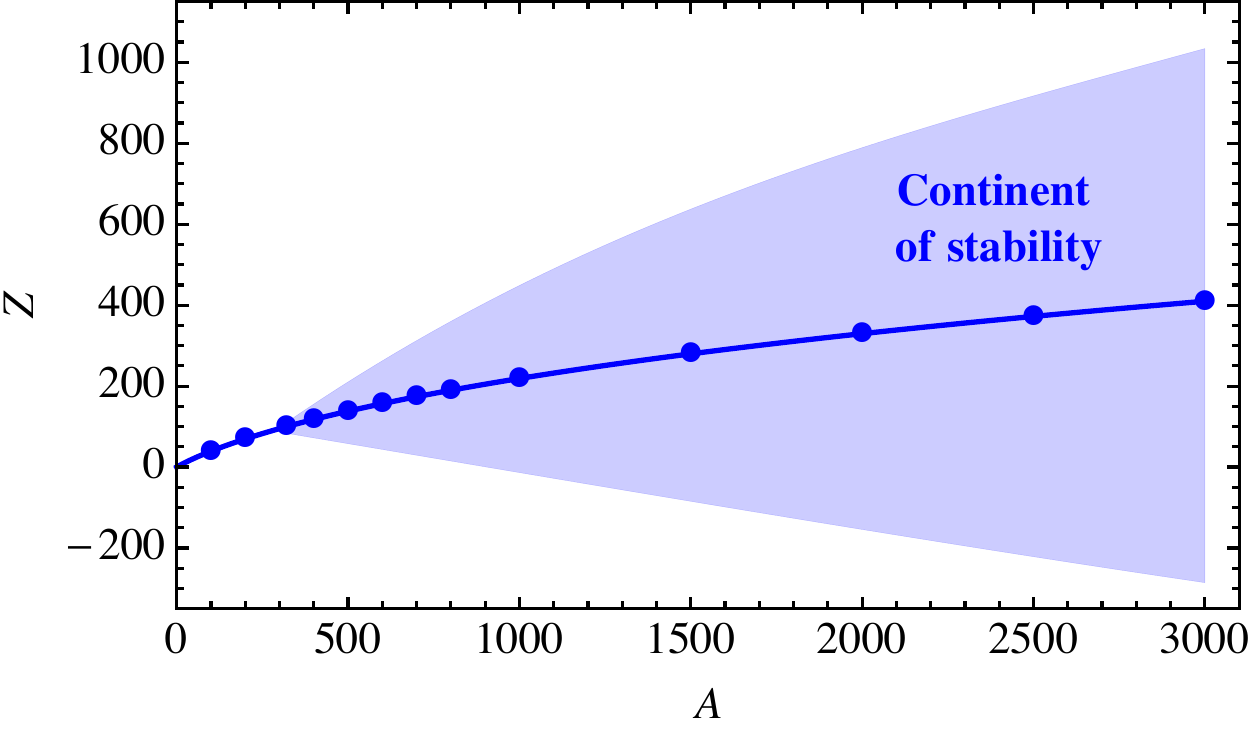}  
 \caption{The electric charge of $ud$QM: full result (blue dots) and the bulk approximation (blue line).}
\label{lowZA}\vspace{-3ex}
\end{figure}

\begin{figure}[t]
\includegraphics[width=8cm]{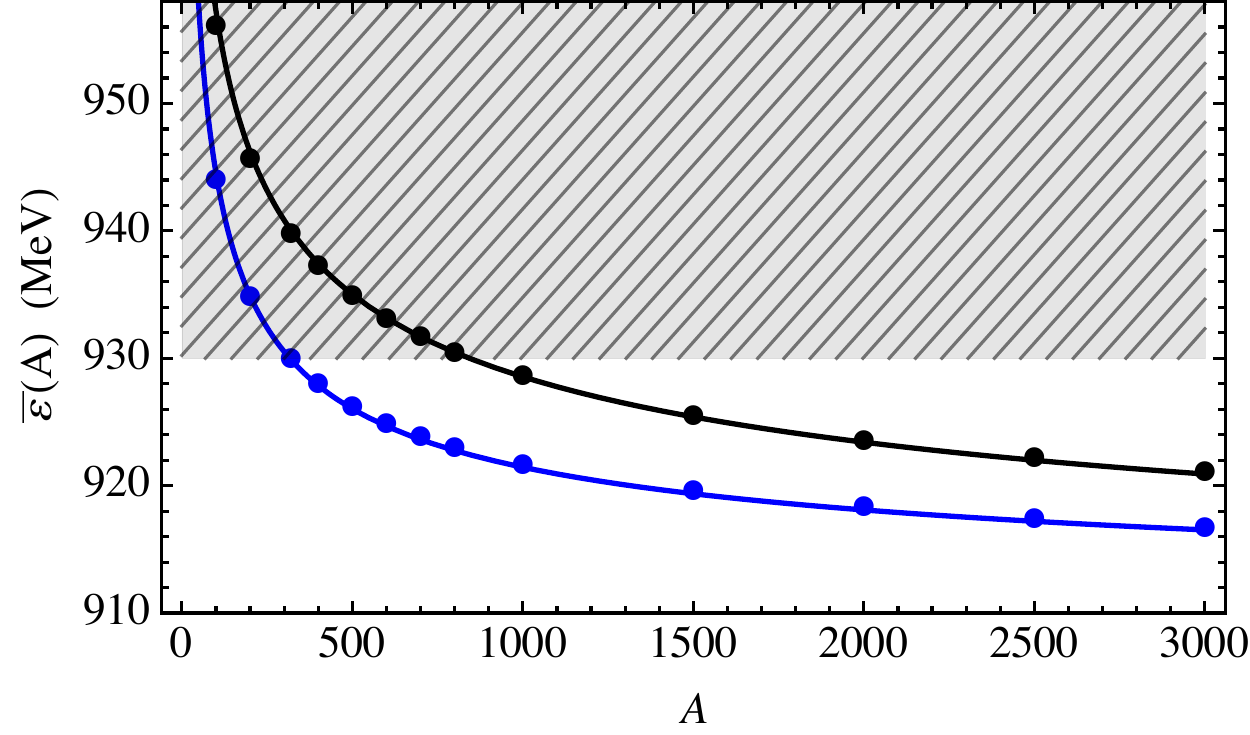}  
 \caption{The minimal energy per baryon $\bar\varepsilon(A)$ for $ud$QM (lower), compared to the charge neutral configuration (upper).}\vspace{-3ex}
\label{lowNmini}
\end{figure}

For the Set 1 benchmark with $m_{udv}=330$~MeV, the numerical solutions of the electric charge and the minimal energy per baryon as functions of $A$ are presented by blue dots in Fig.~\ref{lowZA} and Fig.~\ref{lowNmini} respectively. 
It turns out that the electric charge of $ud$QM can be well estimated by simply minimizing the quark and Coulomb energies of the relativistic $u,\,d$ gas with charge $Z=N_CA(\bar{f}_u(A)-1/3)$, as shown by the blue line in Fig.~\ref{lowZA}. We find $Z \approx \frac{0.86}{\alpha N_C } A^{1/3}$ for large $A$.
The shaded region denotes configurations that are stable against decays into ordinary nuclei. 

Fig.~\ref{lowNmini} shows that the surface effect increases the energy and destabilizes the $ud$QM configuration for $A< A_\textrm{min}$. For Set 1 (Set 2) $A_\textrm{min}\approx 320$ (450) is large enough to prevent normal nuclei from decaying to $ud$QM.
The numerical results of $\bar\varepsilon(A)$ can be well approximated by incorporating a surface tension term $4 \pi R^2 \Sigma$ into the bulk analysis:  $\bar\varepsilon(A)\approx \bar\varepsilon+46\, \Sigma/(\bar{p}_{F}^2A^{1/3})+0.31 \, \alpha \,Z^2\bar{p}_{F}/ A^{4/3}$. Here $\bar\varepsilon$ and $\bar{p}_{F}$ reflect the value of $\bar\chi$ for given $Z$ and $A$. From fits from the two Sets and other examples we find that $\Sigma\approx (91\, \textrm{MeV})^3$, and that it varies less than $\bar\varepsilon$ as displayed in Fig.~\ref{scan}. So as long as $\bar\varepsilon\gtrsim$ 903\,MeV we can expect that $A_\textrm{min}\gtrsim 300$. The surface term dominates the Coulomb term for all $A$, and so the analog of fission that ends the periodic table does not occur for $ud$QM.

\section{Discussion}

If $A_\textrm{min}$ for $ud$QM is close to the lower limit, it raises the hope to produce this new form of stable matter by the fusion of heavy elements. With no strangeness to produce this may be an easier task than producing SQM.
Due to the shape of the curve in Fig.~\ref{lowZA}, there would still be the issue of supplying sufficient neutrons in the reaction to produce $ud$QM, as in the attempts to produce normal superheavy nuclei in the hypothetical ``island of stability'' around $A\approx300$~\cite{Karpov:2017upn}. $ud$QM may instead provide a new ``continent of stability'' as shown in Fig.~\ref{lowZA}, in which the largest values of $Z/A$ are of interest for production and subsequent decay to the most stable configuration. As with SQM, the further injection of neutrons (or heavy ions ~\cite{Farhi:1985ib,PerilloIsaac:1998xm}) can cause the piece of $ud$QM to grow with the release of an indefinite amount of energy \cite{Shaw:1988pc}.  

Neutron stars could convert to $ud$ quark stars. Due to the potential barrier generated by the surface effects, the limiting process for the conversion of a neutron star is the nucleation of a bubble of quark matter initially having the same local flavor composition as the neutron star, via a quantum or thermal tunneling process \cite{Bombaci:2016xuj}. There is then a subsequent weak decay to the stable state, SQM or $ud$QM, as the bubble grows. The barrier for conversion leads to the possibility that there can co-exist both neutron stars and quark stars \cite{Drago:2015cea}. In comparison to SQM, $ud$QM predicts a smaller $\bar{\rho}=4\Delta V_n$ given the same $\bar\varepsilon$. So $ud$ quark stars allow a larger maximum mass, which is of interest for pulsars with $M \sim 2M_\odot$~\cite{Demorest:2010bx, Antoniadis:2013pzd}. The possible superconducting nature of quark matter in stars may have interesting implications, more so for its transport rather than bulk properties~\cite{Weber:2004kj}.  

\begin{acknowledgments}
\noindent\textbf{Acknowledgments. }  

This research is supported in part by the Natural Sciences and Engineering Research Council of Canada. 

\end{acknowledgments}

\end{document}